\documentclass[useAMS]{mn2e}
\usepackage{psfig}

\title[Modulation Doppler Tomography]{Extending emission line Doppler Tomography ; mapping modulated line flux}

\author[D.Steeghs]
{D.Steeghs\\
Harvard-Smithsonian Center for Astrophysics, MS-67, 60 Garden Street,
Cambridge, MA 02138, USA\\
Department of Physics and Astronomy, University of Southampton, Highfield, Southampton S017 1BJ, UK}

\date{Accepted ;
      Received ;
      in original form}

\pagerange{\pageref{firstpage}--\pageref{lastpage}}
\pubyear{2003}

\begin{document}
\maketitle
\label{firstpage}

\begin{abstract}

Emission line Doppler tomography is a powerful tool that resolves the
accretion flow in binaries on micro-arcsecond scales using time-resolved spectroscopy.
I present an extension to Doppler tomography that
relaxes one of its fundamental axioms and permits the mapping of 
time-dependent emission sources. 
Significant variability on the orbital period is a common
characteristic of the emission sources that are observed in
the accretion flows of  cataclysmic variables and X-ray
binaries. Modulation Doppler tomography maps sources varying
harmonically as a function of the orbital period through the
simultaneous reconstruction of three Doppler tomograms. One image  
describes the average flux distribution like in standard tomography, while the two additional images describe the
variable component in terms of its sine and cosine amplitudes. 
%These
%tomograms describe the mean and variable component of the line flux
%distribution in the Doppler coordinate frame.
%
I describe the implementation of such an extension in the form of the maximum entropy based fitting code
MODMAP. Test reconstructions of synthetic data illustrate that the
technique is robust and well constrained. Artifact free reconstructions of complex
emission distributions can be achieved under a wide range of signal to noise levels. An application of the technique is illustrated by mapping
the orbital modulations of the asymmetric accretion disc emission in the dwarf
nova IP Pegasi.

\end{abstract}

\begin{keywords}
binaries: close --- accretion, accretion discs --- novae, cataclysmic
variables --- line: profiles --- techniques: spectroscopic
\end{keywords}

\section{Introduction}

The presence of prominent emission lines in the spectra of
close binaries is one of the tell-tale signs of an accretion
flow. The highly structured and time dependent emission line profiles
observed in cataclysmic variables (CVs) and X-ray binaries harbour a wealth
of information. Since the flow is highly supersonic, the overall line profile shape
is not determined by the local line profile. Instead, it is dominated by the 
dynamics of the flow, thus
providing a direct kinematical signature of the accretion structure
(Smak 1981, Horne \& Marsh 1986).  
This motivated the
development of Doppler tomography (Marsh \& Horne 1988)  as a tool to
recover the accretion flow from the observed line profiles in CVs.
It was recognised that the observed line profile at each orbital phase is a
projection of the accretion flow along the line of sight, and given
sufficient observed projection angles, the profiles can be inverted
to give the line emission distribution in the binary.  Such different
projection angles are naturally provided to us as the binary
components orbit their common centre of mass as a function of the
orbital period permitting reconstructions that resolve the accretion
flow on micro-arcsecond scales.
I refer to Marsh (2002) and Horne (1991) for the basic details concerning
the inversion process from data to Doppler maps, a process which
shares many similarities with medical CAT-scanning procedures.
Since one observes the radial velocities of the emitting material in the
line profiles through Doppler shifts, the choice of working in velocity space is one of the
major advantages of standard Doppler tomography. While not perhaps as
intuitive to interpret as spatial images in the Cartesian X-Y frame,
Doppler tomograms in  $V_x-V_y$ velocity space can be obtained without
 specific prior assumptions concerning the velocity field in the flow as a
function of position. This significantly simplifies the inversion
process and allows the application of Doppler tomography in a variety
of conditions where the nature of the flow is not a-priori known.
Since its conception, Doppler tomography has grown to become a widely used
tool with over 100 papers in the refereed press presenting Doppler
images. For recent reviews and a number of examples of its application
see Boffin, Steeghs \& Cuypers (2002).
Despite its flexibility, Doppler tomography rests on certain
assumptions and approximations. In this paper, I present an extension
to Doppler tomography that relaxes one of its main assumptions, thus further
increasing its applicability. I will start by reviewing the basic
axioms of tomography in Section 2, the implementation of the extension
is described in Sections 3 and 4, and the new code is scrutinised in
Sections 5 and 6.

\section{The axioms of Doppler tomography}

Standard tomography reconstructs the distribution of line emission in
the binary in a velocity coordinate frame. It effectively decomposes
the observed line profiles into discrete emission sources
characterised by their velocity vector in the co-rotating frame of the binary.
I start by reviewing the basic axioms of Doppler tomography as given in Marsh
(2002), albeit in a different order;

\begin{enumerate}
\item the intrinsic line profile width is negligible
\item all velocity vectors co-rotate in the binary
\item motion is mapped parallel to the orbital plane only
\item all points are equally visible at all times
\item the flux from any point is constant in time
\end{enumerate}

\noindent Violations of these assumptions do not imply that Doppler tomography
cannot be performed, or that the maps are meaningless, but these
assumptions must be borne in mind when interpreting Doppler maps. As
such, the tomogram is a representation of the observed data given the
above assumptions and the accuracy and reliability of that
representation is thus intimately tied in with the validity of the above axioms. 

For example, the local line profile is expected to be set by a thermally
broadened profile, amounting to a width of order 10 km/s for typical CV
temperatures. Compared to Doppler velocities of hundreds to thousands
of km/s makes axiom 1 generally satisfied under a wide range of
conditions. However, non-thermal broadening mechanisms may be at work
in parts of the flow, and can in particular affect low velocity
emission. 
Axioms 2 and 3 can be violated in parts of the
flow, but the derived Doppler map should then be seen as
approximations whereby the flow is projected onto the orbital frame of
the binary and the map provides a time averaged representation of the
flow in the co-rotating frame.
Of the five axioms, number four is the most difficult to deal with
since one does not generally know the full geometry of the flow that is
being mapped. After all, the whole point of Doppler mapping is to probe this
 geometry.  In high inclination systems, self-shadowing effects
may play a role, and complicate the analysis of emission line profiles. 

This leads us to the remaining axiom, that is the flux from each point
in the co-rotating frame is assumed constant in time. However,
observations show that the typical emission sources one is mapping do
modulate their flux in time. This can be because the line
source emits an-isotropically (e.g. the emission from density waves and
shocks in an accretion disc) or because axiom 4 is violated and the geometry is
responsible for a strong modulation of the flux (e.g. the buried bright spot
or the irradiated secondary star) or a combination of both.
Doppler tomography can still be used in that
case, since the Doppler map serves to present a time
averaged image of the distribution of line emission, ignoring such
phase-dependent complications. However, one will not
be able to fit the data very well, and the phase dependent information
contained in the observed line profiles is lost. This motivated the
development of an extension to Doppler tomography that allows the line
flux from a given point to vary as a function of time. 
The next Sections concern the implementation of such an extension
which I refer to as modulation Doppler tomography.

\section{Mapping modulated emission}

In conventional Doppler tomography, each line source is characterised
by its inertial velocity vector ${\bf V}=(V_x,V_y)$, where the binary centre of mass is at the
origin, the x-axis points from the accretor to the donor and the
y-axis point in the direction of motion of the donor, as viewed in
the co-rotating frame of the binary. The intensity of the Doppler map
at a given velocity $I(V_x,V_y)$ describes the line flux observed from
the velocity element with centre $(V_x,V_y)$ and
 width $(dV_x,dV_y)$. Each velocity vector follows a sinusoidal
radial velocity curve $V(\phi)$ in the inertial frame of the observer as a function
of the orbital phase $\phi$ given by:

\begin{equation} 
V(\phi) = \gamma - V_x \cos{2\pi\phi} + V_y\sin{2\pi\phi} 
\end{equation}

\noindent where $\gamma$ is the systemic velocity of the binary system
and phase zero is defined to be inferior conjunction of the donor
star with $0 < \phi < 1$. Each line source contributes a constant amount to the observed line
profiles at each phase (axioms 4 and 5), and the observed line profile $F(v_r,\phi)$ is simply a
projection of the radial velocities and intensities of all velocity
vectors considered;

\begin{equation} 
F(v_r,\phi) = \int I(V_x,V_y) \star g(V-v_r) \mbox{ } dV_xdV_y 
\end{equation}

\noindent with $g(V-v_r)$ describing the local line profile intensity at a Doppler
shift of $V-v_r$, which is usually assumed to be a Gaussian profile
with  a width corresponding to the instrumental resolution.
Let us now consider line sources that modulate harmonically in flux as
a function of the orbital phase;

\begin{equation} 
f(\phi) = I_{avg} + I_{cos}\cos{2\pi\phi} + I_{sin}\sin{2\pi\phi} 
\end{equation}

\begin{figure}
\label{spot}
\psfig{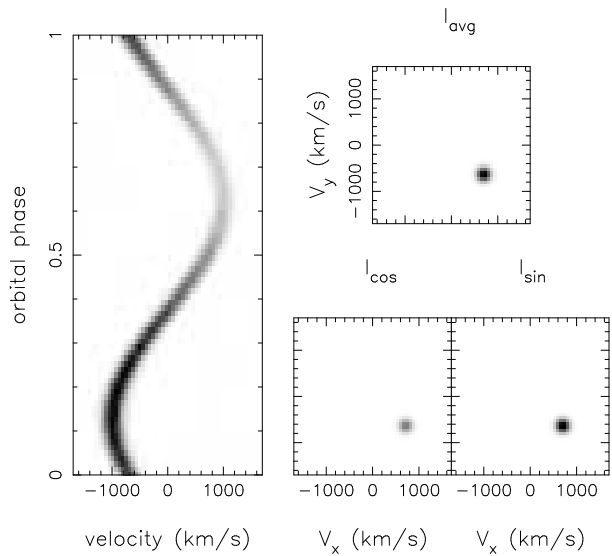}
\caption{The signature of a time dependent emission source is not only
characterised by its radial velocity curve $V(\phi)$, but also its
emission strength along its velocity trajectory as a function of time;
$f(\phi)$. $V(\phi)$ is characterised by its coordinate $(V_x,V_y)$ in
the Doppler coordinate frame (Eqn. 1) whereas $f(\phi)$ is
characterised by the amplitudes $I_{avg}$, $I_{cos}$ and
$I_{sin}$ (Eqn. 3). All modulations on the orbital period can then be described
as the sum of the image values in the three Doppler images.}
\end{figure}

\noindent One then has three parameters to describe the line flux from a
specific velocity vector; the average line flux $I_{avg}$ and the amplitude and
phase of the modulation in terms of $I_{cos}$ and $I_{sin}$. This in contrast to the single parameter
$I(V_x,V_y)$ in Doppler tomography as described above.
In this prescription, we need three Doppler maps describing the
values of $I_{avg}$,$I_{cos}$ and $I_{sin}$ for each velocity
$(V_x,V_y)$ (see Figure 1). 
 The total amplitude of the modulation is then
$\sqrt{I_{cos}^2+I_{sin}^2}$, and its phase is $\tan^{-1}(I_{cos}/I_{sin})$.
In terms of Equation 2, this amounts to replacing $I(V_x,V_y)$ with $I_{avg}(V_x,V_y) + I_{cos}(V_x,V_y)
\cos(2\pi\phi) + I_{sin}(V_x,V_y)\sin(2\pi\phi)$.
This extension is thus straightforward to implement, and does not
fundamentally alter the nature of the projection from data to maps and
vice-versa. Rather then constructing one Doppler map, one now has 
three Doppler maps characterising the nature of the emission across
the co-rotating frame. 
While I could have allowed for a more complicated type of modulation,
either by including more terms to account for non-sinusoidal
modulations, or by letting the period of modulation to be different from
the orbital period, the above implementation provides a simple and
flexible prescription with minimal additional free parameters to fit
the observed data with.

\section{Implementation of MODMAP}

The inversion process from observed data to Doppler maps is normally
done via either the back-projection method in combination with Fourier
filtering (Horne 1991) or maximum entropy regularisation (Marsh \&
Horne 1988). For a comparison of the relative merits of both methods in the light of Doppler tomography see Marsh (2002).
For the implementation of the modulation Doppler tomography code, I
choose to employ the maximum entropy route, since a back-projection
would suffer from serious cross-talk between the three image terms and
produce significant imaging artifacts\footnote{Keith Horne
implemented such an extension as is described here and indeed found
strong artifacts in test reconstructions when using filtered
back-projection (Horne, private communication).}. 
The Fortran code MODMAP uses the MEMSYS algorithm (Skilling \& Bryan
1984) in order to iteratively adjust the Doppler images while achieving a user 
specified $\chi^2$.  The algorithm gradually steers toward the
requested $\chi^2$ value while maximising the entropy of the images in
order to converge to a unique solution. It delivers the simplest image
that can fit to data to the specified level, whereby simplicity is
reflected by the image entropy (see also Horne 1994).
This image entropy ($S$) of an image vector ${\bf I}=I(1,..N)$ is defined relative to a template image of the same dimension, the
default map ${\bf D}$:

\[ S = \sum^{N}_{i=1} (D(i) - I(i) ) + I(i) ln(I(i)/D(i))  \]

\noindent Using such a maximum entropy regularised fitting scheme, the final solution is determined by the requested $\chi^2$ value and the choice of
default.  The entropy is maximised (approaches zero) as ${\bf I}$
approaches ${\bf D}$.
%
%While the choice of an appropriate default is an important
%issue for eclipse mapping, Doppler maps are well constrained by the
%data, and the commonly used scheme is to use an adjusta
%
Since the entropy is only defined for positive image values, but the
cosine and sine amplitudes of the modulation can be both positive or
negative, the Doppler image vector passed to the fitting code
was split into a sequence of five velocity maps.
The first image array represents the average image values
($I_1=I_{avg}$), while the cosine and sine values are represented by two
image arrays each. For those two images, one image reflects positive
amplitudes, the other negative amplitudes such that $I_{cos}=I_2-I_3$
and $I_{sin}=I_4-I_5$. In this way all possible modulations are included
while a positive image vector is maintained under all circumstances
for which an entropy can be defined.

For the default image vector, I experimented with several options,
and the commonly used running default method using a blurred version
of the image vector was implemented because of its good convergence
properties.  In this scheme, the default is recalculated after each
iteration by heavily blurring the current image in velocity space,
providing a smooth default that only contains large scale image structure. In order to minimise structure in the
modulation images $I_2 - I_5$, the defaults for these
were steered to zero after blurring. 
The velocity grid ($N_x \times N_y$) of the Doppler images is user defined, finer grids
requiring significantly more CPU time for each iteration since the
number of pixels in the image vector is $5 \times N_x \times
N_y$. In order to speed up the iterative fitting procedure, an optimal
value was sought for the initial average image ($I_1$) through $\chi^2$
minimisation of a constant scaling factor. The other four images were
started off at zero. The {\sc MODMAP} code would then iterate to a user
specified $\chi^2$ until converged at the maximum entropy solution.

\section{Test reconstructions}

In order to test the reliability of the modulation Doppler tomography
technique, and to explore image artifacts due to both random and systematic
errors, a large set of test data sets were generated. Time-dependent
line profiles were synthesised and passed to the MODMAP code so that
its reconstructions could be compared with the original input data.
In this Section, I discuss a selection of these reconstructions. They
will illustrate the robustness of the image reconstructions, and
explore the possible effects of external systematic errors.

\subsection{Reconstruction 1: Independent spots}

The first test data set concerns the reconstruction of three isolated 
emission spots, each at a different inertial velocity and with different
 modulation characteristics. 
The first spot does not modulate at all, has a FWHM of 100 km/s and
unit strength. The second spot is broader (250 km/s) and modulates
strictly in a cosine fashion with 90\% amplitude, while the third spot
 modulates strictly in sine fashion with an amplitude of 50\%. 
The corresponding Doppler maps for these spots are presented in Figure
2, and test data was generated from these maps. After adding  
random Poisson-type noise, the simulated data was loaded into the
MODMAP code in order to reconstruct the Doppler maps. 
As Figure 2 illustrates, the code reconstructs these
spots perfectly, in terms of their velocity location, overall 
strength and width as well as modulation amplitudes. Most importantly, no
cross talk is present between the modulation images in the sense that
there is no remnant of a pure cosine source in the sine image and vice
versa.

\begin{figure*}
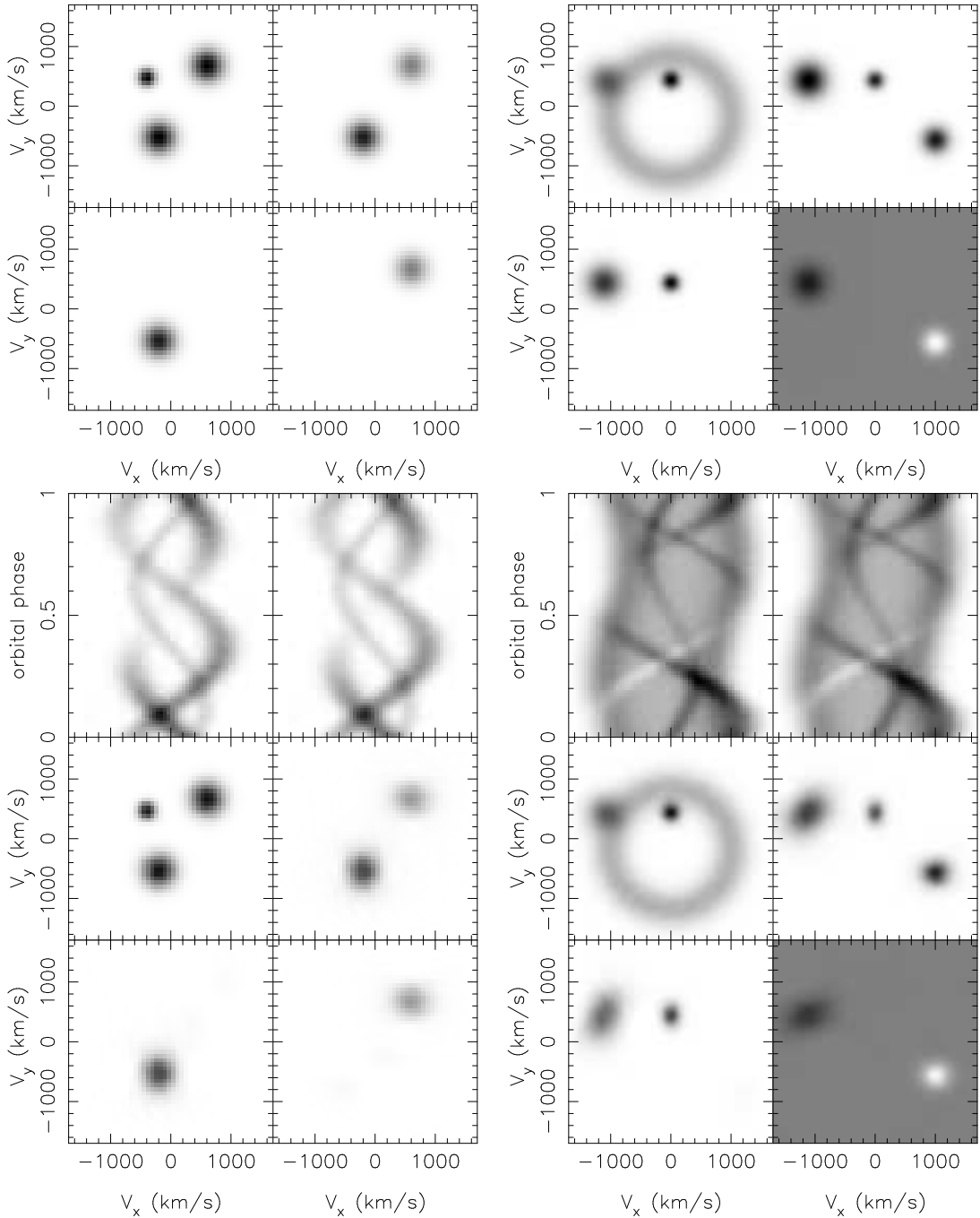

\label{spot}
\begin{tabular}{cc}
\psfig{figure=fig2a.eps,width=7cm} &
\psfig{figure=fig2b.eps,width=7cm}\\
\psfig{figure=fig2c.eps,width=7cm} & 
\psfig{figure=fig2d.eps,width=7cm}\\
\end{tabular}
\caption{Examples of test reconstructions based on synthetic data
sets. Top panels indicate the input Doppler maps, expected to be
recovered given perfect signal to noise and data sampling. The four
maps in each panel correspond to, starting top-left and moving in clockwise fashion; the average
emission, the total modulated emission, the sine amplitude map and the
cosine amplitude map. Below the cosine input maps are the generated line
profiles, after Poisson noise was added. The recovered Doppler images
are plotted in the bottom panels, and the predicted line profiles
based on these maps, used to evaluate the goodness of fit,  are
presented to the right of the input line profiles. The grayscale is
chosen such that black corresponds to the maximum value of
the images, and white to the lowest. In case of negative image values,
zero corresponds to 50\% gray.}
\end{figure*}

\subsection{Reconstruction 2: Disc plus spots}

For the second reconstruction, I chose a more complex and demanding 
emission pattern reminiscent of the common emission sources found in
CVs. 
The first component
was a non-varying accretion disc contribution centred at slightly negative $V_y$
velocities (the velocity of the accretor), represented by a ring in the Doppler coordinates. Then a
narrow spot was added at the typical location of the mass donor star 
($V_x=0$,$V_y=400$ km/s). This spot modulates in a purely cosine fashion
with 50\% amplitude and it is just separated from the disc emission in terms of
its velocity. A second spot representative of a disc hot-spot was
added with a contribution to all three Doppler images (40\% modulation
amplitude in both the sine and cosine terms). Its velocities
overlap with that of the disc emission. Finally, a third spot was
included, again overlapping with the disc emission in terms of
velocity, and this time with a pure sine modulation with a negative
amplitude of 50\%. I refer to Figure 2 for the corresponding input
images and data, and the results from the MODMAP reconstruction.
The code again recovers the original image structure very well. The
modulation amplitudes are reproduced correctly, and the spots are
cleanly separated from the disc emission. Thus even sources that overlap in
velocity but have different modulation characteristics are reproduced
correctly and without artifacts or cross talk.
This holds true even at low signal to noise levels. In Figure 3a we
illustrate the same reconstruction, after degrading the signal to
noise of the input data to only 5. As expected, 
the image structure recovered is a blurred version of
the original, due to the low signal to noise of the data. However, the spot
properties are still reproduced correctly and without cross-talk.
Provided that the modulation is of 
sufficient amplitude to be significant given the signal to noise of
the data, the reconstructions are stable for a wide range of signal to
noise levels. In such cases where low signal is a concern,
 the significance of reconstructed features can be assessed using Monte-Carlo methods (Marsh 2002). An ensemble of
reconstructions should be made using bootstrapped copies of the input data,
in order to quantify the pixel-dependent variance in the Doppler tomograms.

These two test reconstruction served to illustrate the robustness of
the reconstruction method, and to confirm that artifact
free reconstructions can be reconstructed from complex time-dependent
emission sources. 
When reconstructing Doppler maps from real data, two parameters need
to be established; the systemic velocity $\gamma$, and the local line
profile function $g(v)$. The next two reconstructions illustrate the
artifacts that can be introduced by a significant error in the estimates
for these parameters.

\subsection{Reconstruction 3: Incorrect systemic velocity}

\begin{figure*}
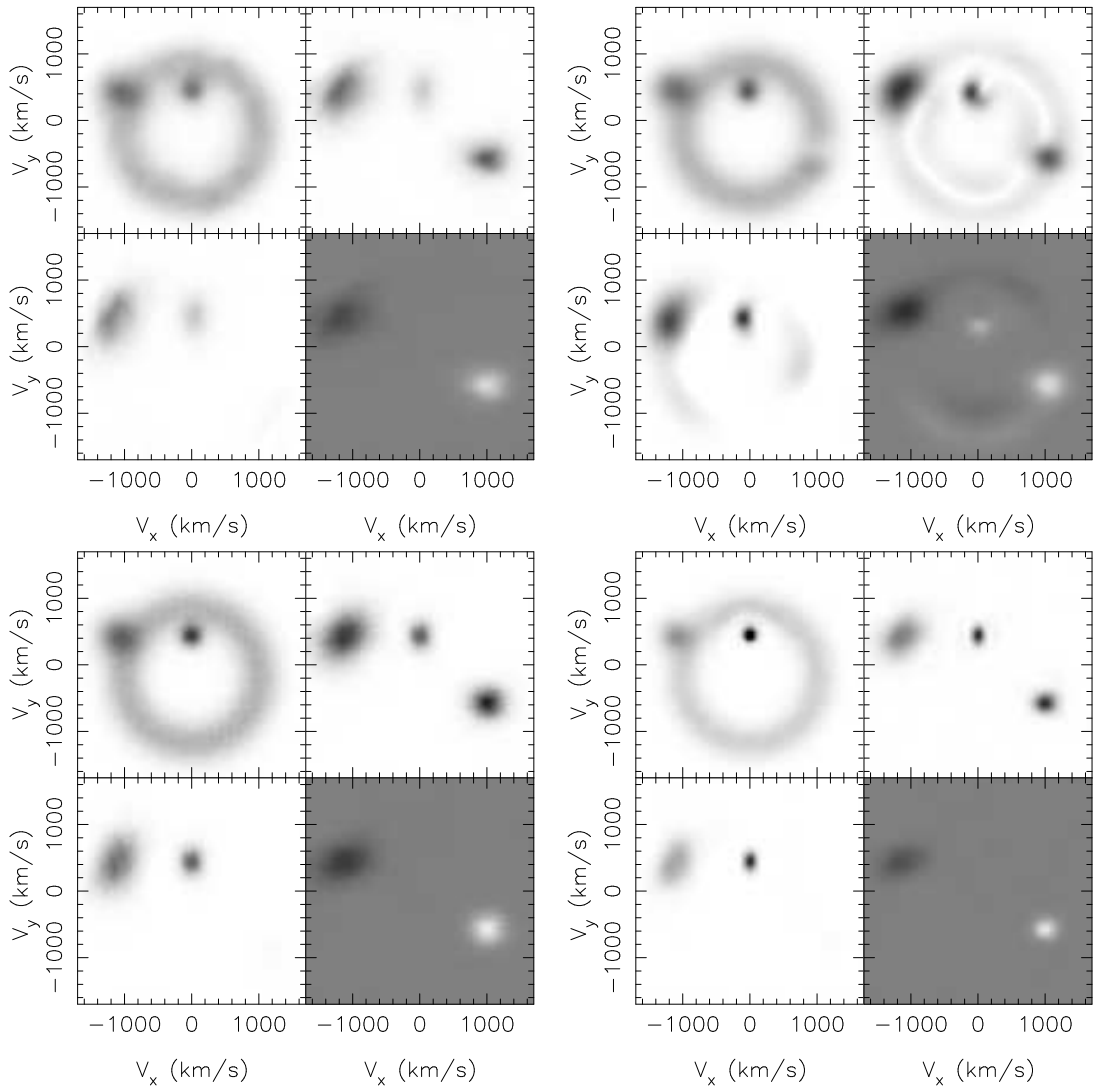

\label{gamma}
\begin{tabular}{cc}
\psfig{figure=fig3a.eps,width=7cm} & 
\psfig{figure=fig3b.eps,width=7cm} \\
 \psfig{figure=fig3c.eps,width=7cm} &
\psfig{figure=fig3d.eps,width=7cm}\\
\end{tabular}
\caption{Test reconstructions of synthetic data based on the same
input maps as presented in Figure 2b. Top left (a) is reconstruction after
degrading the signal to noise of the data to 5. Top right (b) is the
reconstruction achieved while assuming an incorrect systemic
velocity. The bottom plots illustrate the result after
underestimating the local line profile width by a factor of two (c; left),
and overestimating it by the same factor (d; right).}
\end{figure*}

The systemic velocity needs to be known in order
to calculate the correct radial velocity curve (Eqn 1) of a line
source with velocity ($V_x$,$V_y$). If one attempts to reconstruct a data
set with an incorrect gamma, it will be difficult to achieve a good
fit to the data since the code is not tracing the correct velocity curves (see also Marsh \& Horne 1988). Figure 3b illustrates the
reconstructed Doppler images  using the same data set that was employed
in the previous test reconstruction, but now reconstructing it with a
systemic velocity that is incorrect by 60 km/s. Clear artifacts are
present in all the maps, and the code, as expected, 
is unable to achieve a reduced $\chi^2 \sim 1$. 
Thus, an incorrectly assumed gamma can lead to
significant artifacts in the modulation maps, more so than in
the average map. The total modulation flux map illustrates that the
(non-varying) disc in particular leaves artifacts in the modulation images. 

For many systems, an estimate of the systemic velocity is known or can
be measured from the line profiles. The uncertainty in $\gamma$ only
becomes an issue if its error approaches or exceeds the local line
profile width. In those cases where the data resolution is
very high, or independent $\gamma$ estimates are not available, the
achieved $\chi^2$ value can be employed as a self-consistent test of
the assumed systemic velocity. Since an incorrect $\gamma$ leads to image artifacts
and poor data fits, one can perform a series of Doppler map
reconstructions while trying to minimise $\chi^2$ as a function of
$\gamma$ in order to identify the correct systemic velocity.  Thus, provided the potential effects of an incorrect gamma are
borne in mind, its impact on the image reconstructions can usually be verified 
self-consistently, and can be used to determine the systemic velocity.

\subsection{Reconstruction 4: Error introduced by incorrect local line
profile}

The local line profile is assumed to be a Gaussian with its width
adjusted to the instrumental resolution of the data.
 In Figure 3c, I illustrate the reconstructed Doppler
maps after underestimating the local line width by a factor of
two. While the maps show weak fine structure due to this oversampling
effect, it still recovers the correct image structure.
 On the other hand, overestimating the local line width by the same
factor leads to a sharpening of the recovered maps (Figure 3d); the
reconstructed spot sizes are now too small in the Doppler images and
the convergence is very slow. This is the result of the code attempting to
reproduce the observed width of the spots with an assumed local line
profile that is too broad. This can only be achieved by making the spot
widths smaller in the maps. The effect of a forced error in $g(v)$
affects all the maps in the same way, and does not degrade the ability
to recover modulated emission sources. It is clearly better to caution on the
side of oversampling by underestimating the local line width to avoid
artificial sharpening of the tomograms. 
Thus provided the structures that one is
imaging are sampled properly, reconstructions are robust against
modest local line width variations.

\section{Application to IP Pegasi}

As a real world example of the use of modulation mapping, I applied
MODMAP to existing data of the eclipsing dwarf nova IP Pegasi during outburst.  This data
together with conventional Doppler maps were presented in 
Harlaftis et al. (1999) and revealed strong spiral arms in the accretion disc.
Figure 4 plots the reconstructed modulation Doppler maps for the HeII
4686\AA~emission of IP Pegasi. A much better reconstruction of the
observed data was achieved using the MODMAP code compared to the
conventional Doppler reconstruction, indicative of significant orbital variability.
The average map corresponds in detail to the maps presented in Harlaftis et
al. (1999), with a disc dominated by two spiral arms and weak
secondary star emission. Most
of the modulated flux is dominated by the sine term, with modulation amplitudes of up to 20\%. Again, the variable disc emission is concentrated in two arms. However,
these arms do not correspond to the location of the spirals in detail, but
only overlap partially. Instead, the modulated disc flux is at
slightly larger velocities (i.e. closer in) and rotated anti-clockwise
in terms of phase. It thus appears that most of the modulation is not
due to anisotropic radiation from the spiral arms directly, but
variable emission from the disc regions just inside. This could be due to
geometric shielding of those disc areas by the vertically extended
spiral structures. The phasing of the modulation supports such an
interpretation; the emission from the disc just inside the closest arm
is weaker compared to that of the disc inside the opposite arm and
both modulate in anti-phase. This is expected if the spiral arms are
vertically extended, shielding the disc area immediately behind them
from the observer at high inclinations (IP Pegasi is an eclipsing
system after all).  As always, one must bear in mind that although the
modulation mapping code is able to show that disc flux is modulated,
significant geometrical shielding violates the basic assumptions of
Doppler tomography (axiom 4). 

As well as modulated emission from the disc, the maps also show that the
secondary star and an extended low velocity region are time
dependent emission sites. 
This reconstruction illustrates the ability of MODMAP to identify
emission sites in the binary that modulate as a function of the
orbital phase and characterise their modulation in terms of its phase
and amplitude. It achieves better fits to the data, and aids in
establishing the nature of the line emitting sources. A more 
extensive and quantitative analysis of reconstructions using MODMAP on existing data
sets will be presented in future papers.

\begin{figure}
\label{ippeg}
\psfig{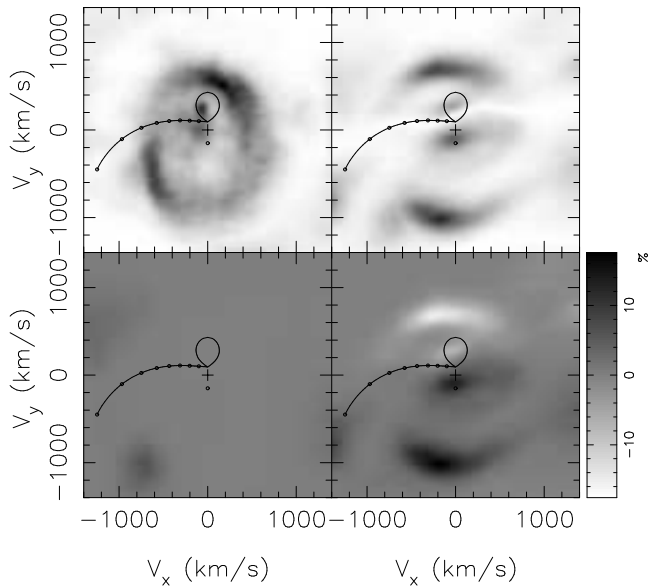}
\caption{MODMAP reconstructions of the HeII emission from IP Pegasi
during outburst (data from Harlaftis et al. 1999). The emission from
the asymmetric accretion disc carrying a two armed spiral modulates
significantly. The grayscale wedge denotes the fractional modulation amplitudes
in the sine and cosine images.}
\end{figure}

\section{conclusions}

I presented an extension to emission line Doppler tomography that
permits the mapping of time dependent emission sources. It relaxes 
one of the five fundamental axioms of conventional Doppler tomography, and
maps harmonically varying modulations on the orbital period. 
Since significant variability on the orbital period is a common
characteristic of the typical emission sources in CVs and X-ray
binaries, this prescription permits a more detailed reconstruction of
the nature of the accretion flow.   Intrinsic variability not
correlated with the orbit is however not described by this extension.
Instead of a single Doppler map describing the time averaged emission
distribution in the co-rotating frame of the target, modulation Doppler
tomography describes the emission through three Doppler tomograms. One
describes the average flux distribution just like in standard Doppler
tomography, while the two additional tomograms describe the
variable component in terms of its sine and cosine amplitudes. Thus
for each location in the $V_x-V_y$ Doppler coordinate frame, the
mean flux and the amplitude and phase of any variable flux
contribution is given by the three image values at that velocity.

Test reconstructions using the maximum entropy based fitting code
MODMAP in conjunction with synthetic data show that the technique is robust, and artifact free
reconstructions of complex emission distributions can be achieved
under a wide range of signal to noise levels. 
The effects of systematic uncertainties in the assumed systemic
velocity and local line profile width are similar to those seen in
conventional tomography.

There are several reasons to expect anisotropic emission from the
accretion flow. Horne (1995) explores the signature of anisotropic
turbulence on the emission pattern of a turbulent accretion
disc. Anisotropy in the turbulence Mach matrix also leads to
anisotropic line emission from the disc. The prominent spiral
structures seen in dwarf nova accretion discs are also likely to emit
an-isotropically (e.g. Steeghs \& Stehle 1999) as was illustrated in
Section 6.
The highly variable emission from the disk-stream impact region
(e.g. Spruit \& Rutten 1998) is another prime example of orbitally
modulated emission. In Morales-Rueda et al. (2003), the results from
modeling the bright spot emission in GP Com using MODMAP are
presented.
Finally, the irradiated secondary
star is a common emission line source present in the line
profiles (e.g. Marsh \& Horne 1990). While the emission from the 
Roche-lobe shaped donor is better
mapped with Roche-tomography techniques (Watson \& Dhillon 2002),
modulation Doppler tomography is able to reproduce the basic
properties of the emission. This permits much better fits to the data,
making a strong secondary star contribution less problematic for disc
mapping experiments (Steeghs 2002).

Modulation Doppler tomography is a robust and well constrained
extension that may assist the interpretation of both existing and
future data sets of accreting binaries. As with all tomographic
imaging techniques, it nevertheless still relies on several
assumptions that need to be borne in mind when interpreting any image
reconstructions (Section 2). Its main advantage over conventional Doppler
tomography is to describe emission sources that vary harmonically as
a function of the orbital phase. Its added flexibility should find a
wide range of uses, in particular in conjunction with high signal to
noise data.

%Modulation Doppler tomography code is one particular example of the potential
%of extending our suite of tomography tools. The development was
%motivated by the rather common presence of time dependent emission
%sources. 

\section*{Acknowledgments}

The author acknowledges support from a Smithsonian Astrophysical
Observatory Clay Fellowship and was supported by a PPARC Fellowship
when this work was initiated. Keith Horne and
Tom Marsh are thanked for many useful discussions on tomography and
maximum entropy related topics. Thanks to Emilios Harlaftis for his
role in securing the IP Pegasi outburst data.

\label{lastpage}
\end{document}